\documentclass[12pt,preprint]{aastex}

\shorttitle{}

\shortauthors{Liu et al.}

\begin{document}

\title{Sun-to-Earth Characteristics of Two Coronal Mass Ejections Interacting near 1 AU: 
Formation of a Complex Ejecta and Generation of a Two-Step Geomagnetic Storm}   

\author{Ying D. Liu\altaffilmark{1}, Zhongwei Yang\altaffilmark{1}, Rui Wang\altaffilmark{1}, 
Janet G. Luhmann\altaffilmark{2}, John D. Richardson\altaffilmark{3}, and No\'{e} Lugaz\altaffilmark{4}}

\altaffiltext{1}{State Key Laboratory of Space Weather, National 
Space Science Center, Chinese Academy of Sciences, Beijing 100190, China; 
liuxying@spaceweather.ac.cn}

\altaffiltext{2}{Space Sciences Laboratory, University of California, 
Berkeley, CA 94720, USA}

\altaffiltext{3}{Kavli Institute for Astrophysics and Space Research, 
Massachusetts Institute of Technology, Cambridge, MA 02139, USA}

\altaffiltext{4}{Space Science Center, University of New Hampshire,
Durham, NH 03824, USA}

\begin{abstract}

On 2012 September 30 - October 1 the Earth underwent a two-step geomagnetic storm. We examine the Sun-to-Earth characteristics of the coronal mass ejections (CMEs) responsible for the geomagnetic storm with combined heliospheric imaging and in situ observations. The first CME, which occurred on 2012 September 25, is a slow event and shows an acceleration followed by a nearly invariant speed in the whole Sun-Earth space. The second event, launched from the Sun on 2012 September 27, exhibits a quick acceleration, then a rapid deceleration and finally a nearly constant speed, a typical Sun-to-Earth propagation profile for fast CMEs \citep{liu13}. These two CMEs interacted near 1 AU as predicted by the heliospheric imaging observations and formed a complex ejecta observed at Wind, with a shock inside that enhanced the pre-existing southward magnetic field. Reconstruction of the complex ejecta with the in situ data indicates an overall left-handed flux rope-like configuration, with an embedded concave-outward shock front, a maximum magnetic field strength deviating from the flux rope axis and convex-outward field lines ahead of the shock. While the reconstruction results are consistent with the picture of CME-CME interactions, a magnetic cloud-like structure without clear signs of CME interactions \citep{lugaz14} is anticipated when the merging process is finished. 

\end{abstract}


\section{Introduction}

Solar storms, known as coronal mass ejections (CMEs), are massive expulsions of plasma and magnetic flux from the solar corona into interplanetary space. Although CMEs have been recognized as drivers of major space weather effects, it is not clear how their Sun-to-Earth characteristics are connected with the generation of geomagnetic storms. This missing link is a major obstacle to making progress in space weather forecasting. 

Recent studies have focused on the interaction between different CMEs, which is expected to be a frequent phenomenon in interplanetary space especially near solar maximum. CME-CME interactions were first discovered by \citet{gopalswamy01} using SOHO/LASCO imaging observations and enhanced broadband radio emissions. In situ signatures resulting from interactions between CMEs, defined as complex ejecta by \citet{burlaga01, burlaga02}, in general do not show well ordered magnetic fields at 1 AU. Complex ejecta can be very geoeffective given their prolonged periods. They have been considered as a trigger of two-step geomagnetic storms \citep{farrugia04, farrugia06}, in addition to the sheath-ejecta mechanism \citep{tsurutani88}. A new class of complex ejecta was proposed by \citet{lugaz14} based on MHD simulations. This new type of complex ejecta also arises from CME-CME interactions, but unlike typical complex ejecta their magnetic fields manifest a smooth rotation over an extended duration. Clearly, CME-CME interactions complicate the interpretation of in situ signatures observed at 1 AU and make forecasting of geomagnetic activity more difficult. 

With wide-angle heliospheric imaging observations from the Solar Terrestrial Relations Observatory \citep[STEREO;][]{kaiser08}, it is now possible to follow details of the CME interaction process continuously and connect CME interaction features in images with in situ signatures at 1 AU. The first attempt in this regard was performed by \citet{liu12} on the successive CMEs from 2010 July 30 to August 1. This study reveals that a shock overtaking the preceding ejecta plays an important role in the momentum and energy transfer between the interacting CMEs and the shock structure/strength can be modified on a global scale by the interaction, confirming previous MHD simulations \citep[e.g.,][]{vandas97, schmidt04, lugaz05, xiong06}. The study, together with other examinations of the same events \citep{harrison12, juan12, temmer12, mostl12, webb13, zhang14}, indicates that the CME interaction is an inelastic collision process \citep[also see simulations by][]{schmidt04, lugaz05}. This is probably a nature determined by a magnetized plasma that is prone to magnetic reconnection \citep{gosling08}. A more recent study combining remote-sensing and in situ observations finds that interactions between consecutive CMEs can result in a superstorm in interplanetary space with an exceptionally minor deceleration and extremely enhanced ejecta magnetic fields \citep{liu14}. All these indicate the crucial importance of CME-CME interactions for both basic plasma physics and space weather.   

On 2012 September 30 the Space Weather Prediction Center at the National Oceanic and Atmospheric Administration (NOAA) initially predicted a G1 (minor) geomagnetic storm, which was later revised as a G3 (strong) level. This event is likely a two-step geomagnetic storm. We identify the solar and interplanetary source conditions responsible for this two-step geomagnetic storm, which could be otherwise lost during transit to the Earth without the aid of wide-angle imaging observations. The focus of this Letter is to trace the formation of a new type of complex ejecta \citep[recently proposed by][]{lugaz14} from an observational point of view and investigate how the complex ejecta is connected with the generation of the two-step geomagnetic storm. Compared with the complex ejecta discussed by \citet{burlaga01, burlaga02} and \citet{lugaz14}, the present case was in a relatively early stage of the merging process between two CMEs. We can thus see the merging as it occurs, in particular with the advantage of having simultaneous heliospheric imaging and in situ observations. This is key to understanding the formation of complex ejecta as well as CME-CME interactions. Connection between the complex ejecta and the two-step geomagnetic storm also gives crucial information on how CME Sun-to-Earth characteristics result in geomagnetic storm activity. 

\section{Observations and Analysis}

Figure~1 shows the positions of STEREO A and B with respect to the Earth and the CMEs of interest as viewed from the two spacecraft. STEREO A and B were 125.2$^{\circ}$ west and 118.1$^{\circ}$ east of the Earth at a distance of 0.97 and 1.07 AU from the Sun, respectively. The first CME (CME1) occurred on 2012 September 25 as a streamer blowout (Fig.~1b, c) with a peak speed of only about 430 km s$^{-1}$. A possible associated feature on the Sun is a C1.1 flare from NOAA AR 11575 (N08$^{\circ}$W04$^{\circ}$) that peaked around 09:43 UT on September 25. CME1 deviated a little southward, so the Earth probably encountered its flank. The second CME (CME2) has a maximum speed of about 1200 km s$^{-1}$ and is associated with a long-duration C3.7 flare from NOAA AR 11577 (N09$^{\circ}$W31$^{\circ}$) that peaked at about 23:57 UT on September 27. The two CMEs are likely to interact since their propagation directions are close to each other (Fig.~1a). Composite images of the heliospheric imagers (HI1 and HI2) aboard STEREO A indicate that CME2 was chasing CME1 from behind (Fig.~1f). The interaction is expected to occur at a distance of about 1 AU from the Sun given the CME speeds and launch times on the Sun. 

The time-elongation maps, produced by stacking the running-difference images within a slit along the ecliptic plane \citep{sheeley08, davies09, liu10a, liu10b}, are displayed in Figure~2. In STEREO A observations CME2 reached the elongation of the Earth earlier than in STEREO B images, so it must propagate west of the Sun-Earth line. CME1 faded before getting into the FOV of HI2 for STEREO B, but judging from the trend of the track we also expect a propagation direction west of the Sun-Earth line for CME1. Tracks from CME1 and CME2 seem to finally intersect as can be seen in the time-elongation maps from STEREO A, which again suggests that CME1 and CME2 would interact. Comparing the tracks to the observed shock arrival times at the Earth helps establish the connections of CME1 and CME2 with their near-Earth solar wind signatures. 

We can also determine CME Sun-to-Earth kinematics using a triangulation technique based on the time-elongation maps originally proposed by \citet{liu10a}. The technique assumes a relatively compact CME structure simultaneously seen by the two spacecraft. Later, the assumption of a spherical front attached to the Sun for CME geometry is incorporated into the triangulation technique \citep{lugaz10, liu10b}. These are essentially the same triangulation concept under different CME geometry assumptions, so they are called triangulation with F$\beta$ and HM approximations respectively \citep[see more discussions in][]{liu13, liu10b}. The same expressions have been developed for this triangulation concept by \citet{davies13} using a self-similar model for which the F$\beta$ and HM geometries are limiting cases. The triangulation technique has proved to be a useful tool for tracking CMEs and connecting imaging observations with in situ signatures \citep[e.g.,][]{liu10a, liu10b, liu11, liu12, liu13, liu14, mostl10, lugaz10, harrison12, temmer12, davies13, mishra13}.

We apply the triangulation technique using elongation angles extracted along the front edges of the leading tracks in the time-elongation maps. The resulting CME kinematics in the ecliptic are shown in Figure~3. Both of the two approximations give propagation angles west of the Sun-Earth line, consistent with the solar source longitudes and expectations from the time-elongation maps. The propagation angles from triangulation with the HM approximation are about 1.4 - 1.9 times those obtained from triangulation with the F$\beta$ assumption. The distances and speeds from the two approximations show no essential differences for small elongations, but for large elongations the F$\beta$ approach gives an apparent acceleration partly because of the non-optimal observation situation of the two spacecraft from behind the Sun \citep[see detailed discussions in][]{liu13}. Therefore, only distances and speeds derived from the HM assumption are given in Figure~3. The Sun-to-Earth propagation of CME2, a typical fast event, shows three phases: a quick acceleration, then a rapid deceleration and finally a nearly constant speed. The above results are similar to what has been found before \citep{liu13}. CME1 is a slow event and only has an acceleration followed by a nearly invariant speed. A possible interpretation is that CME1 is first brought up to about the ambient solar wind speed and then co-moves with the solar wind \citep[e.g.,][]{gosling96, sheeley99, lindsay99, gopalswamy00}. Part of the acceleration could be attributed to the forward drag by the ambient solar wind. A similar speed profile was found for another slow, streamer blowout event \citep[a so-called ``stealth" CME;][]{robbrecht09} by \citet{rollett12} based on a single spacecraft analysis. We tentatively suggest that this Sun-to-Earth propagation profile is typical for slow CMEs and complements the finding of \citet{liu13} for fast ones. 

The predicted arrival time at the Earth based on the distances is about 00:43 UT on September 30 for CME1 and 04:16 UT on September 30 for CME2. The predicted speeds at the Earth are about 380 km s$^{-1}$ and 688 km s$^{-1}$, respectively. We have used $r\cos\beta$ and $v\cos\beta$ in estimating the arrival time and speed at the Earth \citep{mostl11, liu13}, where $r$ is the distance from the Sun, $v$ the speed, and $\beta$ the propagation angle with respect to the Sun-Earth line. The distance profiles of the two CMEs cross around 08:02 UT on September 30 at about 1.07 AU from the Sun. The interaction between CME1 and CME2, however, is likely to commence before 1.07 AU given their nonzero thickness in the radial direction.

The in situ signatures at Wind are plotted in Figure~4. Two shocks, presumably driven by CME1 and CME2 separately, passed Wind at 10:16 and 22:19 UT on September 30, which are about 9.5 and 18 hours later than predicted respectively. The average speeds in the regions behind the shocks are about 305 and 400 km s$^{-1}$, somewhat lower than predicted by the triangulation technique \citep[for explanations see][]{liu13}. The individual CME boundaries cannot be distinguished within the complex ejecta, whose interval is identified based on the discontinuities in the density, temperature and magnetic field. The proton temperature inside the complex ejecta is not depressed as usual for a CME at 1 AU, probably due to interactions between CME1 and CME2. The speed profile is not declining monotonically across the interval, indicating that the merging of the two CMEs is still in process. The southward component of the magnetic field is about $-5$ nT in the first part of the interval but sharply drops to about $-20$ nT after the second shock (S2). A possible interpretation is that S2 was plowing through CME1 and compressing the magnetic field and pre-existing southward component within CME1. Behind S2 the magnetic field shows a rotation over an extended period (see the $B_T$ and $B_N$ components). It is conceivable that, after S2 has exited from the complex ejecta (i.e., when the merging process is complete), the complex ejecta will exhibit a simply declining speed profile, an enhanced magnetic field strength and a southward-to-northward rotation of the magnetic field. These characteristics are similar to those of a magnetic cloud \citep[MC;][]{burlaga81} with a SEN configuration according to the classification of \citet{bothmer98}, except that the proton temperature is not depressed. 

The $D_{\rm st}$ profile in Figure~4 indicates a two-step geomagnetic storm sequence with a global minimum of $-119$ nT. Following a sudden commencement caused by the impact of S1, the first dip in the $D_{\rm st}$ index is produced by the southward magnetic field component in the leading part of the complex ejecta. The $D_{\rm st}$ index appears to recover after the first dip, but this ``recovery" actually results from contamination by the magnetopause current owing to the second strike by S2. Soon the ring current is intensified by the sudden increase in the southward magnetic field component, which leads to the second dip in the $D_{\rm st}$ value. We model the $D_{\rm st}$ index using an empirical formula based on the solar wind measurements \citep{om00}. The simulated $D_{\rm st}$ profile agrees with actual $D_{\rm st}$ measurements, except that the global minimum is underestimated. Application of another empirical formula \citep{burton75} gives a deeper second dip, but also underestimates the actual minimum as well as the whole values during the recovery phase (not shown here). 

Figure~5 shows the cross section of the complex ejecta reconstructed from the in situ data (the shaded interval in Fig.~4) using a Grad-Shafranov (GS) technique \citep{hau99, hu02}, which has been validated by well separated multi-spacecraft measurements \citep{liu08, mostl09}. The reconstruction gives an axis elevation angle of about $13^{\circ}$ and azimuthal angle of about $263^{\circ}$ in RTN coordinates. The overall magnetic field configuration is left-handed, as can be seen from the transverse fields along the spacecraft trajectory. An inspection of the magnetic field orientations along the spacecraft trajectory also reveals that, as Wind moves along $x$ in the flux-rope frame, it would see a $B_R$ component that is largely negative, a $B_T$ that is also largely negative (since the flux-rope axis is almost opposite to \textbf{T}), and a $B_N$ that is first negative and then positive. These expectations are consistent with the in situ measurements in Figure~4, indicating validity of the GS reconstruction. A striking feature in the cross section is that the maximum magnetic field strength and peak axial field do not overlap, which is presumably a result of interactions between CME1 and CME2. Another prominent feature is the concave-outward shock front visible from the color shading and change in the magnetic field lines. The shock location is also consistent with the magnetic field strength profile along the spacecraft trajectory, although it is smoothed by the GS integration scheme. The convex-outward field lines ahead of the shock front probably correspond to the flank of CME1 encountered by the Earth (see Fig.~1). These results agree with the picture of a complex ejecta with an embedded shock formed by interactions between CME1 and CME2, as revealed by combined wide-angle imaging and in situ observations. 

It is surprising that the complex ejecta can be reconstructed by the GS method despite the CME interaction and presence of a penetrating shock. The interaction between the two CMEs is likely such that a local translational symmetry is formed during the interaction process. The shock profile is smoothed by the integration scheme, so the quasi-static assumption of the GS technique could be satisfied. Again, the complex ejecta will have a structure resembling a single flux rope after the shock has exited. This emerging complex ejecta is different from the traditional ones defined by \citet{burlaga01, burlaga02}, which in general have disordered magnetic fields. Its formation from CME merging, however, does conform to the concept of complex ejecta. Although the duration of the complex ejecta is not as long as discussed by \citet{lugaz14}, the idea of an MC-like structure resulting from CME-CME interactions is consistent with their proposed new type of complex ejecta based on MHD simulations.

\section{Conclusions}

We have investigated how CME Sun-to-Earth characteristics lead to the formation of a complex ejecta and the generation of a two-step geomagnetic storm, using coordinated wide-angle imaging observations and in situ measurements. Two CMEs, which occurred on 2012 September 25 and 27, have Sun-to-Earth propagation profiles characteristic of slow and fast events respectively: an acceleration followed by a nearly invariant speed, as opposed to a quick acceleration, then a rapid deceleration and finally a nearly constant speed. They began to interact near 1 AU and produced a complex ejecta with a rotating magnetic field and an embedded shock observed at Wind. The shock driven by the second CME was plowing through the first CME at 1 AU and compressing the pre-existing southward magnetic field component, which gave rise to the two-step geomagnetic storm. The $D_{\rm st}$ index shows an apparent recovery after the first dip, which actually foreshadows a more severe geomagnetic storm level. The reconstructed cross section of the complex ejecta using the GS technique reveals non-overlapping maximum magnetic field strength and peak axial field, an embedded concave-outward shock front and convex-outward field lines ahead of the shock, consistent with the CME-CME interaction scenario. After the shock has exited, the complex ejecta will exhibit an MC-like structure without clear signs of CME interactions. This is essentially the new type of complex ejecta recently proposed based on MHD simulations \citep{lugaz14}. 

\acknowledgments The research was supported by the Recruitment Program of Global Experts of China, NSFC under grant 41374173 and the Specialized Research Fund for State Key Laboratories of China. We acknowledge the use of data from STEREO and Wind and the $D_{\rm st}$ index from WDC in Kyoto.

\clearpage

\begin{figure}
\centerline{\includegraphics[width=21pc]{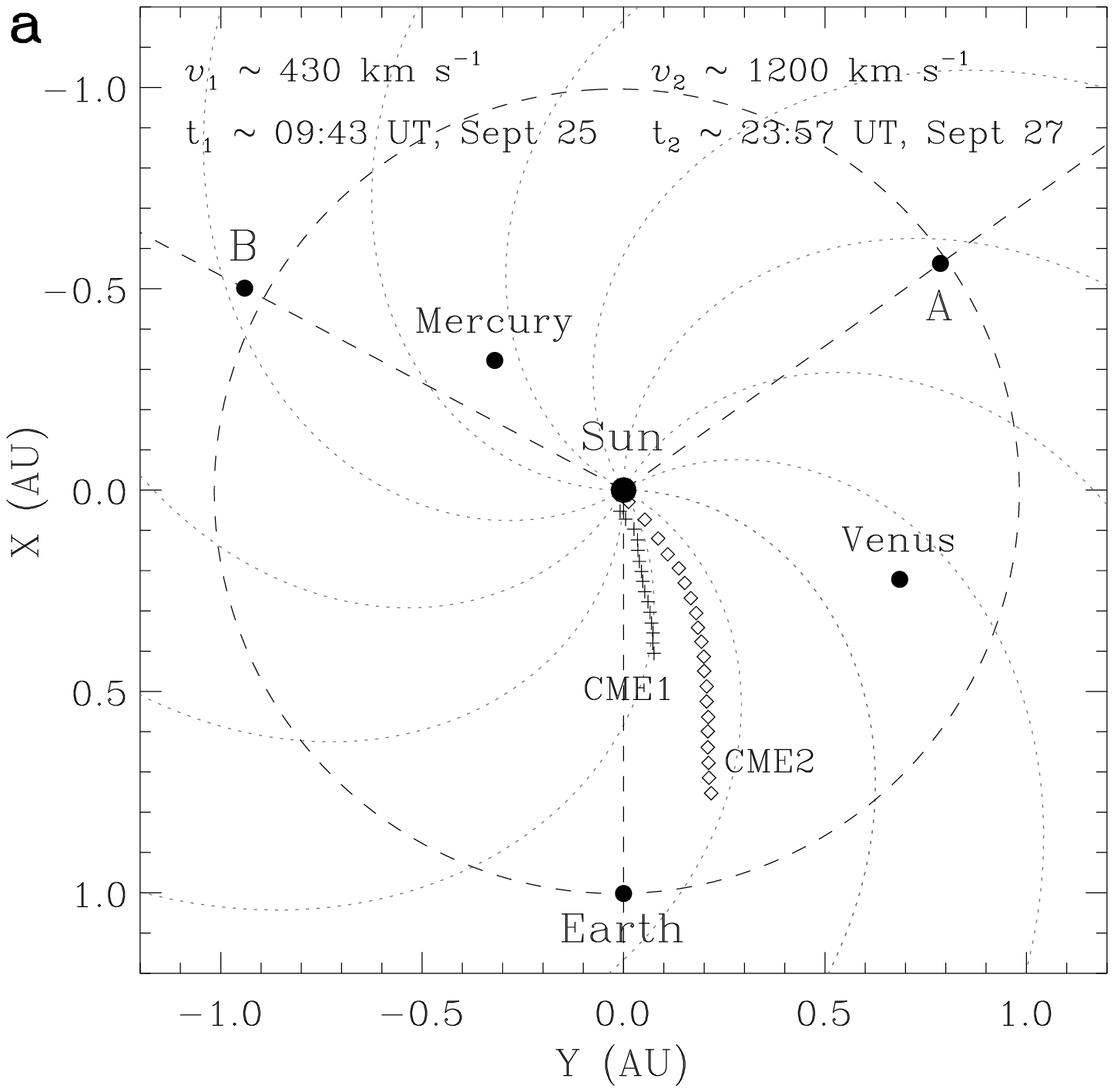}}
\vspace{0.6pc}
\centerline{\includegraphics[width=36pc]{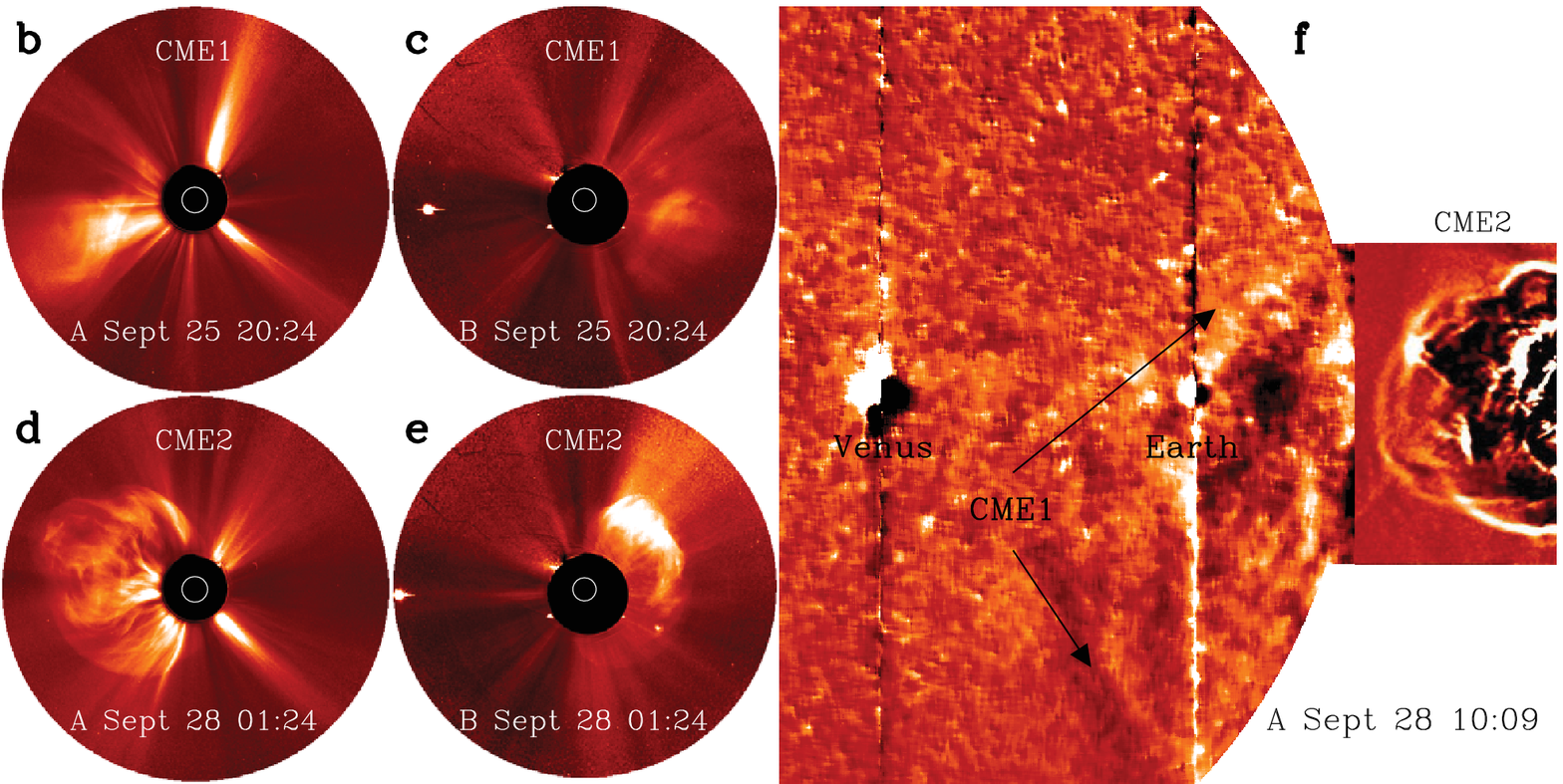}}
\caption{(a) Positions of the spacecraft and planets in the ecliptic plane on 2012 September 28. The dashed circle indicates the orbit of the Earth and the dotted lines show the spiral interplanetary magnetic fields. The trajectories of CME1 and CME2, which are obtained by a triangulation technique assuming the CME front as a spherical structure attached to the Sun, are marked by crosses and diamonds respectively. The estimated CME peak speed and launch time on the Sun are also given. (b-e) COR2 images of the two CMEs viewed from STEREO A (left) and B (right) near simultaneously. (f) Composite image of HI1 and HI2 from STEREO A. Two animations associated with this figure are available in the online journal.}
\end{figure}

\clearpage

\begin{figure}
\epsscale{0.8} \plotone{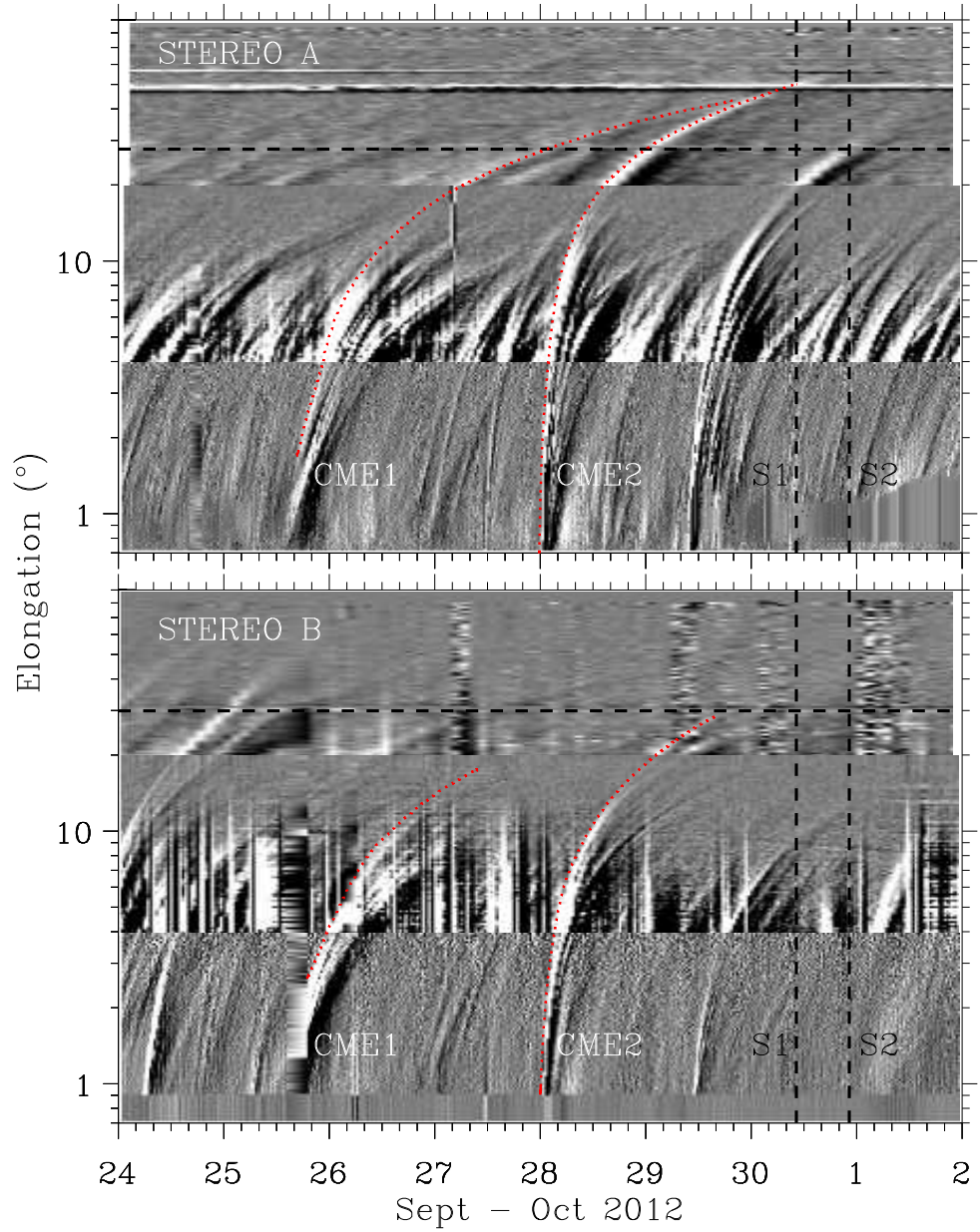} 
\caption{Time-elongation maps constructed from running-difference images of COR2, HI1 and HI2 along the ecliptic for STEREO A (upper) and B (lower). The red dotted curves indicate the front edges of the CME leading tracks. The vertical dashed lines mark the observed shock arrival times at the Earth, and the horizontal dashed line denotes the elongation angle of the Earth.}
\end{figure}

\clearpage

\begin{figure}
\epsscale{0.75} \plotone{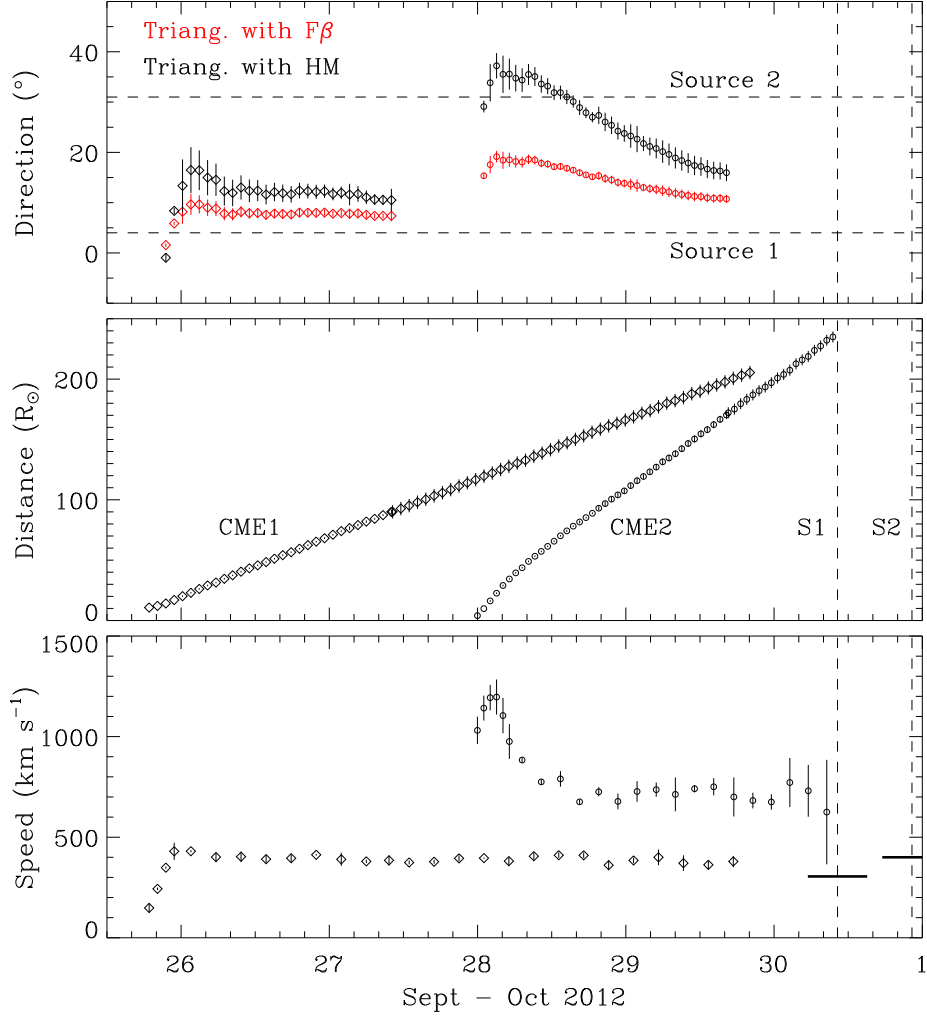} 
\caption{Propagation direction, radial distance and speed of the leading edges of CME1 (diamonds) and CME2 (circles). The longitudes of the CME source locations on the Sun are indicated by the horizontal dashed lines in the top panel. The vertical dashed lines mark the observed shock arrival times at the Earth, and the short horizontal lines in the bottom panel show the corresponding observed average solar wind speeds behind the shocks. The propagation angles are derived from triangulation with F$\beta$ and HM approximations, respectively. For the distance, only values obtained from triangulation with HM are plotted. Elongation measurements of CME1 (CME2) after 10:03 UT on September 27 (16:18 UT on September 29) are available only from STEREO A, so the distances thereafter are calculated from STEREO A observations assuming a propagation direction of 10.5$^{\circ}$ (15.5$^{\circ}$) west of the Sun-Earth line. The speeds are computed from adjacent distances using a numerical differentiation technique. The values are then binned except for the acceleration period, so shown after the acceleration phase are averages and standard deviations within the bins \citep{liu13}.}
\end{figure}

\clearpage

\begin{figure}
\epsscale{0.75} \plotone{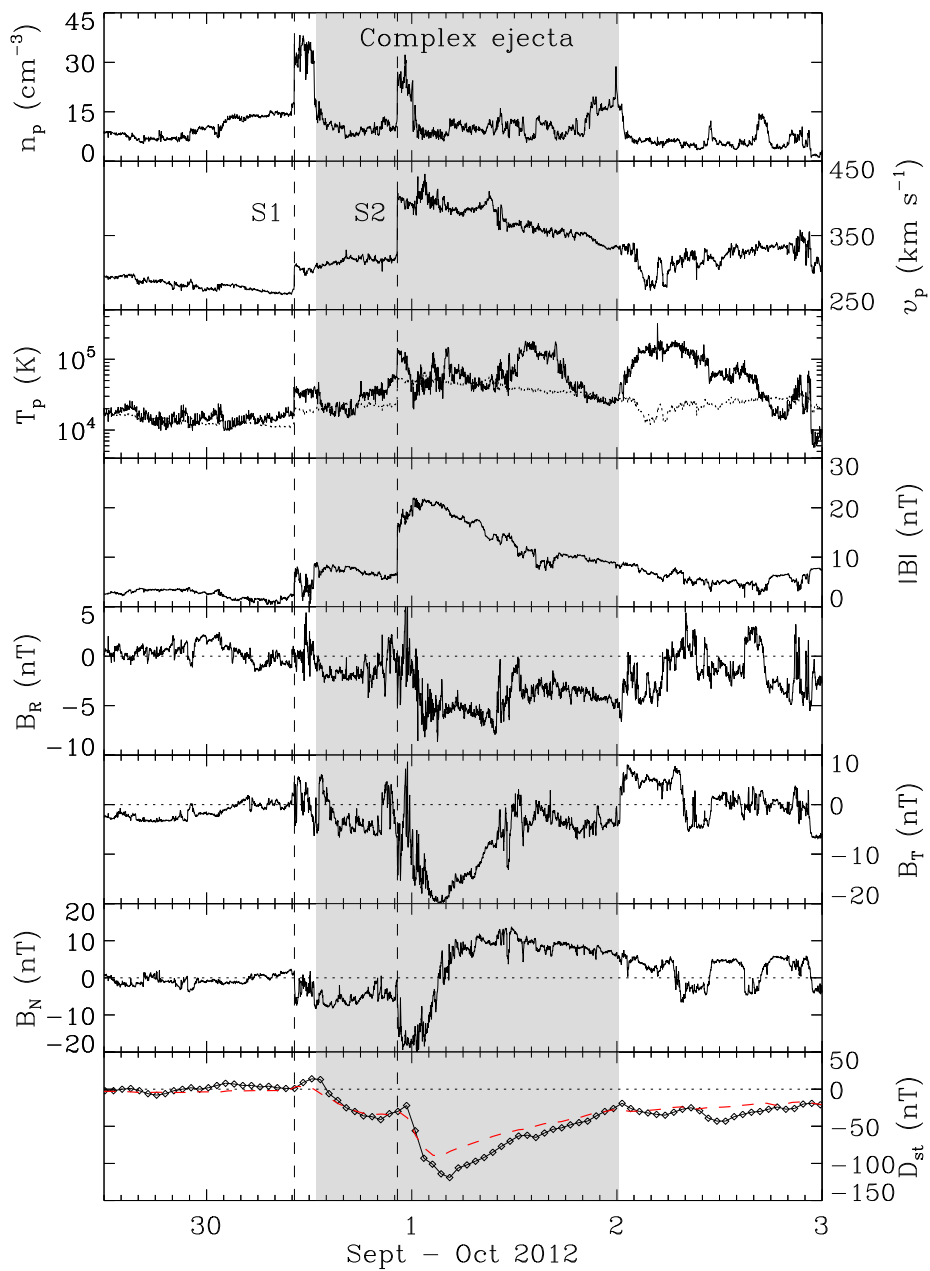} 
\caption{Solar wind parameters observed at Wind. From top to bottom, the panels show the proton density, bulk speed, proton temperature, magnetic field strength and components, and $D_{\rm st}$ index respectively. The dotted curve in the third panel denotes the expected proton temperature from the observed speed. The red curve in the bottom panel represents $D_{\rm st}$ values estimated using the formula of \citet{om00}. The shaded region indicates the interval of the complex ejecta, and the vertical dashed lines mark the associated shocks.}
\end{figure}

\clearpage

\begin{figure}
\epsscale{1.0} \plotone{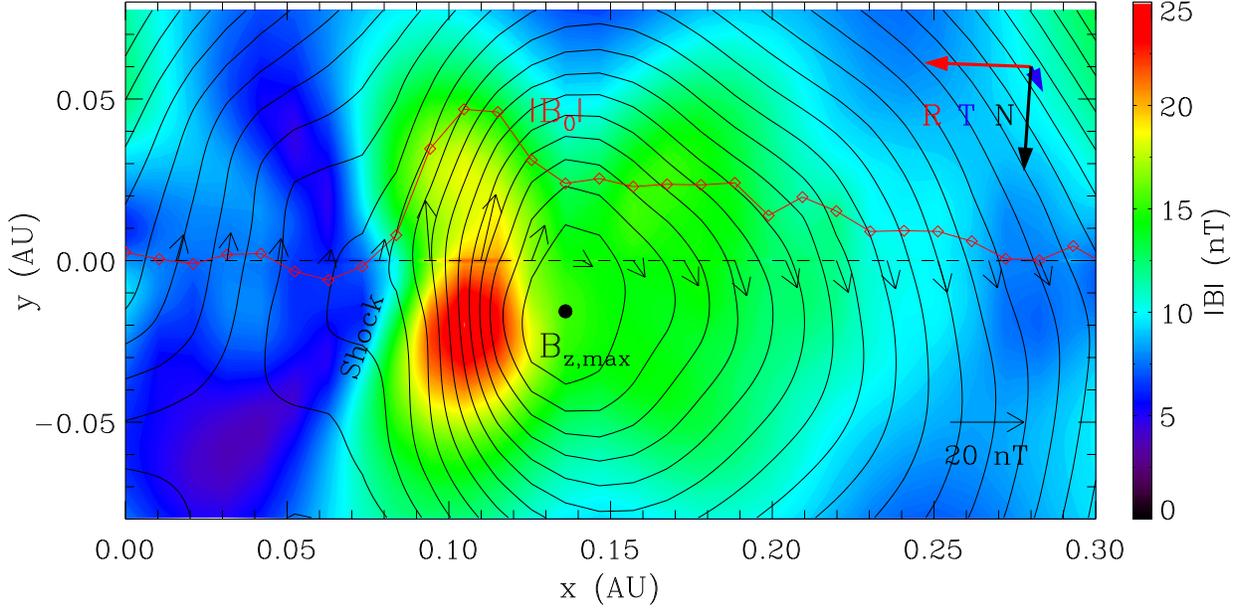} 
\caption{Reconstructed cross section of the complex ejecta. Black contours are the distribution of the vector potential (magnetic field lines), and the color shading shows the value of the magnetic field strength. The location of the maximum axial field is indicated by the black dot. The dashed line marks the trajectory of the Wind spacecraft. The thin black arrows denote the direction and magnitude of the observed magnetic fields projected onto the cross section, and the thick colored arrows represent the projected RTN directions. Overlaid on the cross section is the magnetic field strength along the spacecraft trajectory (red curve; in arbitrary units).}
\end{figure}

\end{document}